\documentclass{rsproca_new}

\usepackage{pstool,booktabs,overpic,multirow,rotating}

\usepackage{tikz}
\usetikzlibrary{intersections,calc,shapes.arrows,shapes.geometric,plotmarks}

\begin{document}

\title{Wave energy absorption by a submerged air bag connected to a rigid float}

\author{A. Kurniawan$^{1,2}$, J. R. Chaplin$^{3}$, M. R. Hann$^{1}$, D. M. Greaves$^{1}$, F. J. M. Farley$^{3}$}

\address{$^{1}$School of Marine Science and Engineering, Plymouth University, Drake Circus, Plymouth, PL4 8AA, UK\\
$^{2}$Department of Civil Engineering, Aalborg University, Thomas Manns Vej 23, 9220 Aalborg, Denmark\\
$^{3}$Faculty of Engineering and the Environment, University of Southampton, Highfield, Southampton, SO17 1BJ, UK}

\subject{xxxxx, xxxxx, xxxx}


\corres{A. Kurniawan\\
\email{aku@civil.aau.dk}}

\begin{abstract}
A new wave energy device features a submerged ballasted air bag connected at the top to a rigid float. 
Under wave action, the bag expands and contracts, creating a reciprocating air flow through a turbine between the bag and another volume housed within the float.
Laboratory measurements are generally in good agreement with numerical predictions.
Both show that the trajectory of possible
combinations of pressure and elevation at which the device is in static equilibrium takes the shape of an S. 
This {means} that statically the device can have three different draughts, and correspondingly three different bag shapes, for the same pressure. 
The 
behaviour in waves depends on where the mean pressure-elevation condition is on the static trajectory. 
The captured power is highest for a mean condition on the middle section.
\end{abstract}

\begin{fmtext}
\section{Introduction}

The idea of using flexible bags for wave energy extraction stretches back to the 1970s~\cite{French1977,French1979}. 
Under wave action, expansion and contraction of the bag create a reciprocating air flow into and out of another volume, driving a turbine in between.

Recently, this concept has been revisited and novel variations have emerged with one thing in common: an axisymmetric bag in the form of a fabric encased within an array of meridional tendons. 
In one variation, the bag is floating and is ballasted such that it pierces the free surface~\cite{Kurniawan2016,Kurniawan2017}, and in another, the bag is fixed at its bottom~\cite{Kurniawan2016b}.

\end{fmtext}

\maketitle

\begin{figure}
\centering
\includegraphics[trim = 0mm 2mm 0mm 2mm, clip, scale=.9]{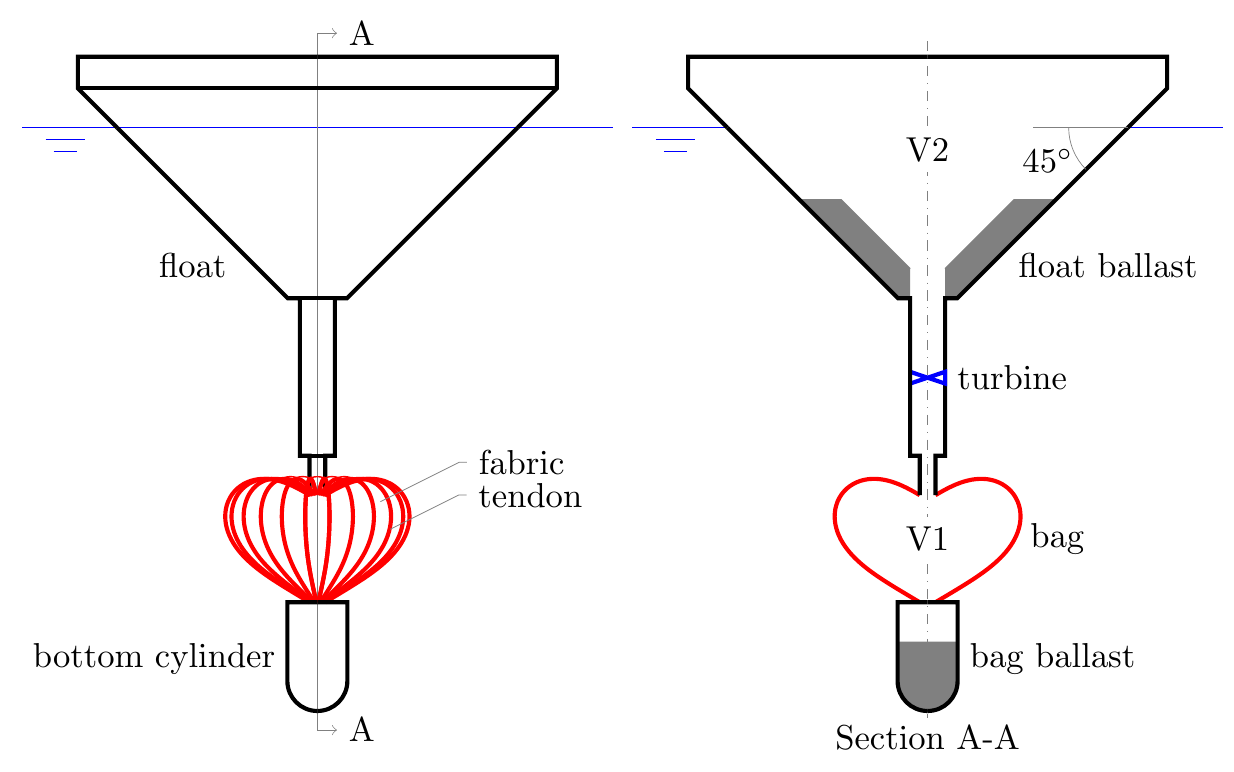}
\caption{Sketch of the device with its main parts identified.}
\label{sq3device}
\end{figure}

This paper considers another variation of the device~\cite{Farley2016patent}, where the bag is fully submerged and connected to a rigid float at the top and to a weight at the bottom (see figure~\ref{sq3device}). 
The motivation was to introduce a double-peaked response, thus broadening the power absorption bandwidth.
The float is herein chosen to be conical in shape.
The two air volumes either side of the turbine, denoted as V1 and V2, are all contained within the device.

A numerical approach calculates the equilibrium shapes of the bag in still water for different bag pressures.
Unlike the bags in the other two device variations~\cite{Kurniawan2017,Kurniawan2016b}, the bag in the present device can have a non-{horizontal} tangent  at the top. 
We predict three different equilibrium shapes for the same pressure, and this is confirmed by physical experiments.
Plotting the elevation of a point on the device against the bag pressure as the amount of air in the bag is slowly varied yields a trajectory in the shape of an S.

To predict the device's response in waves, a linear frequency-domain model  is developed following the approach used for the device with a floating bag~\cite{Kurniawan2017}.
{The effects of air compressibility are included through the linearised isentropic air pressure-density relation, as in an oscillating-water-column device~\cite{SarmentoFalcao1985,MalmoReitan1985,Jefferys1986}}.
Compared to the device with a floating bag~\cite{Kurniawan2017}, the present device has one extra degree of freedom in the motion of the float. 

The response of the device is found to vary depending on where the mean pressure-elevation condition is on the static trajectory. 
The absorbed power is found to be highest when the mean condition is on the middle section {of the trajectory}.
Contrary to what was expected, the absorbed power is not always double-peaked.

\section{Theory}

\subsection{Static calculations}

When the bag is inflated, the fabric forms lobes between the tendons. 
This keeps the tension in the fabric to the minimum, while the tendons carry most of the load.
For simplicity we assume that all tension is carried by the tendons and neglect the volume of the lobes.
This approximation is exact in the limit of infinitely many tendons~\cite{Taylor1919}.
In addition, we assume that the tendons are inextensible and that the bag is weightless.

\begin{figure}
\centering
\includegraphics[trim = 0mm 2mm 0mm 2mm, clip, scale=.9]{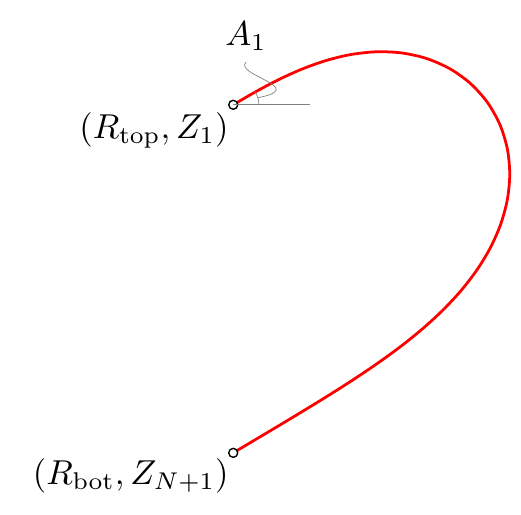}
\caption{Sketch of a single tendon. $R$ is measured from the vertical axis of the device. $Z$ is measured from the mean water line.
{The node angle (shown for $A_1$) is defined as sketched.}}
\label{tendonsketch}
\end{figure}

Due to symmetry, the shape of the bag is defined by the profile of just a single tendon (see figure~\ref{tendonsketch}). 
For a given tendon length as well as top and bottom radii of the bag $R_\mathrm{top}$ and $R_\mathrm{bot}$, the profile of the tendon in still water is determined uniquely by for example the elevation and {angle} (relative to horizontal) at the top of the bag $Z_1$ and $A_1$, and the bag pressure (excluding atmospheric) $P$. 

A method of calculation has been described in~\cite{Kurniawan2017}, which involves discretising the tendon into $N$ arc elements of uniform length $h$ (the total number of nodes is $N+1$).
The element's radius of curvature $\rho_n$ is calculated element-wise from the top of the bag to the bottom, by satisfying the force balance normal to the element:
\begin{equation}
2 \pi h R_{n+0.5} (P + \rho g Z_{n+0.5}) = \frac{T h}{  \rho_n} =  T (A_n - A_{n+1}), \quad n = 1, \dots, N . \label{bagstatic}
\end{equation}
Here, $T$ is the sum of static tension in all tendons,  $\rho$ is the water density, {$g$ is the acceleration due to gravity}, $(R_{n+0.5}, Z_{n+0.5})$ is the radius and elevation at midpoint of element $n$, while $A_n$ is the  {angle} at node $n$.
As $T$ is not known beforehand, an iterative procedure adjusts $T$ so that the difference between the calculated $R_{N+1}$ and the specified $R_\mathrm{bot}$ is less than some small tolerance. 
A similar approach has been used to obtain the profile of inflatable membrane dams under various static loading conditions~\cite{Harrison1970}. 

For a given combination of float's and bottom cylinder's weights, any admissible bag shape must in addition satisfy any two of the following force equilibria: (1) the force equilibrium on the device as a whole, (2) that on the float, and (3) that on the bottom cylinder considered separately.
On the device as a whole, we have
\begin{equation}
M_\mathrm{float} + M_\mathrm{bag} + M_\mathrm{cyl} = \rho (V_\mathrm{float} + V_\mathrm{bag} + V_\mathrm{cyl}) , \label{devicewholestatic}
\end{equation}
where $M_\mathrm{float}$, $M_\mathrm{bag}$, and $M_\mathrm{cyl}$ are the masses of the float, bag, and bottom cylinder, respectively, while $V_\mathrm{float}$, $V_\mathrm{bag}$, and $V_\mathrm{cyl}$ are the corresponding submerged volumes.
On the float, 
\begin{equation}
(M_\mathrm{float}  - \rho V_\mathrm{float}) g = T \sin{A_1} + \pi R_\mathrm{top}^2 (\rho g Z_1 + P) , \label{floatstatic}
\end{equation}
where $A_1$ and $Z_1$  are the  {angle} and elevation at the top of the bag.
On the bottom cylinder,
\begin{equation}
(M_\mathrm{cyl}  - \rho  V_\mathrm{cyl})  g =  - T \sin{A_{N+1}} - \pi R_\mathrm{bot}^2 ( \rho g Z_{N+1} + P) , \label{botcylstatic}
\end{equation}
where $A_{N+1}$ and $Z_{N+1}$  are the  {angle} and elevation at the bottom of the bag.

\subsection{Linear frequency-domain model}

To obtain the response of the device under harmonic excitations, a linear frequency-domain model is developed following the approach described in~\cite{Kurniawan2017}.
Only axisymmetric motions are considered. 
Furthermore, we assume that all displacements are so small that linear theory is valid and that all responses oscillate with the same frequency as the excitation frequency. 
The time-dependent part of any oscillatory quantity can thus be expressed as the real part of the product of a complex amplitude and $\mathrm{e}^{\mathrm{i}\omega t}$, where $\omega$ is the oscillation frequency. 

The approach consists of first expanding the static equations~\eqref{bagstatic},~\eqref{floatstatic}, and~\eqref{botcylstatic} and keeping only terms up to the first order. 
The expanded equations include the hydrodynamic forces. 
Subtracting the static equations from the expanded equations yields a set of linear dynamic equations for the float, bag, and bottom cylinder. 

The boundary conditions at the top and bottom of the bag must be satisfied as well.
The top of the bag does not move radially 
and its vertical motion is the same as that of the float.
Likewise, the bottom of the bag does not move radially 
and its vertical motion is equal to that of the bottom cylinder. 

Finally, by utilising the various relationships outlined in~\cite{Kurniawan2017}, it is possible to express the equations in terms of complex amplitudes of the radial displacements $r_n$ of the tendon element midpoints, the heave displacement of the bottom cylinder $\xi_3$, the relative displacement between the float and the bottom cylinder $\xi_7$, and the dynamic tension $\tau$ in the tendons, as the unknowns.
The final equations can be written in a matrix form as
\begin{equation}
\mathbf{Y} \mathbf{x} = \mathbf{F}, \quad \text{with } \mathbf{x} = \begin{pmatrix} r_1 & \cdots & r_N & \xi_3 & \xi_7 & \tau \end{pmatrix}^T.
\end{equation}
The hydrodynamic forces, which appear in $\mathbf{F}$, if the device is excited by incident waves, and in $\mathbf{Y}$, are computed using a panel method~\cite{WAMIT7.2}. 

Upon solving this linear system of equations, we can obtain the mean captured power from 
\begin{equation}
\mathcal{P} = \frac{C}{ 2 \rho_{\text{air}}} |p_1 - p_2|^2 ,
\end{equation}
where $p_1$ and $p_2$ are the pressure amplitudes in V1 and V2, which can be derived from the volumetric change of the bag {(details are given in~\cite{Kurniawan2017})}, and $C$ is a {real} coefficient that relates the air mass flow through the power take-off (PTO) to the pressure difference across it:
\begin{equation}
\mathrm{i}\omega m_2 = -\mathrm{i}\omega m_1 = C (p_1 - p_2) . \label{C}
\end{equation}
{Note that due to air compressibility the volume flow through the PTO is less than the rate of change of the bag's volume, and there is a phase difference between the two, 
a point discussed in~\cite{MalmoReitan1985} and further elaborated in~\cite{FalcaoHenriques2014}.}

\section{Experiments}

Small-scale model tests were carried out in the 35 m $\times$ 15.5 m wave basin at Plymouth University, with a water depth of 3 m. 

The model bag was constructed by joining two circular unreinforced polyurethane films at the perimeter.
The tendons, of 3-mm diameter Polyester-covered Vectran cores, ran through guides attached to the fabric.
The bag had 12 tendons and each tendon was 0.58 m long.
The tendons terminated both at the top and bottom of the bag at a radius of 0.02 m. 

\begin{figure}
\centering
\includegraphics[trim = 0mm 0mm 0mm 2mm, clip, scale=.9]{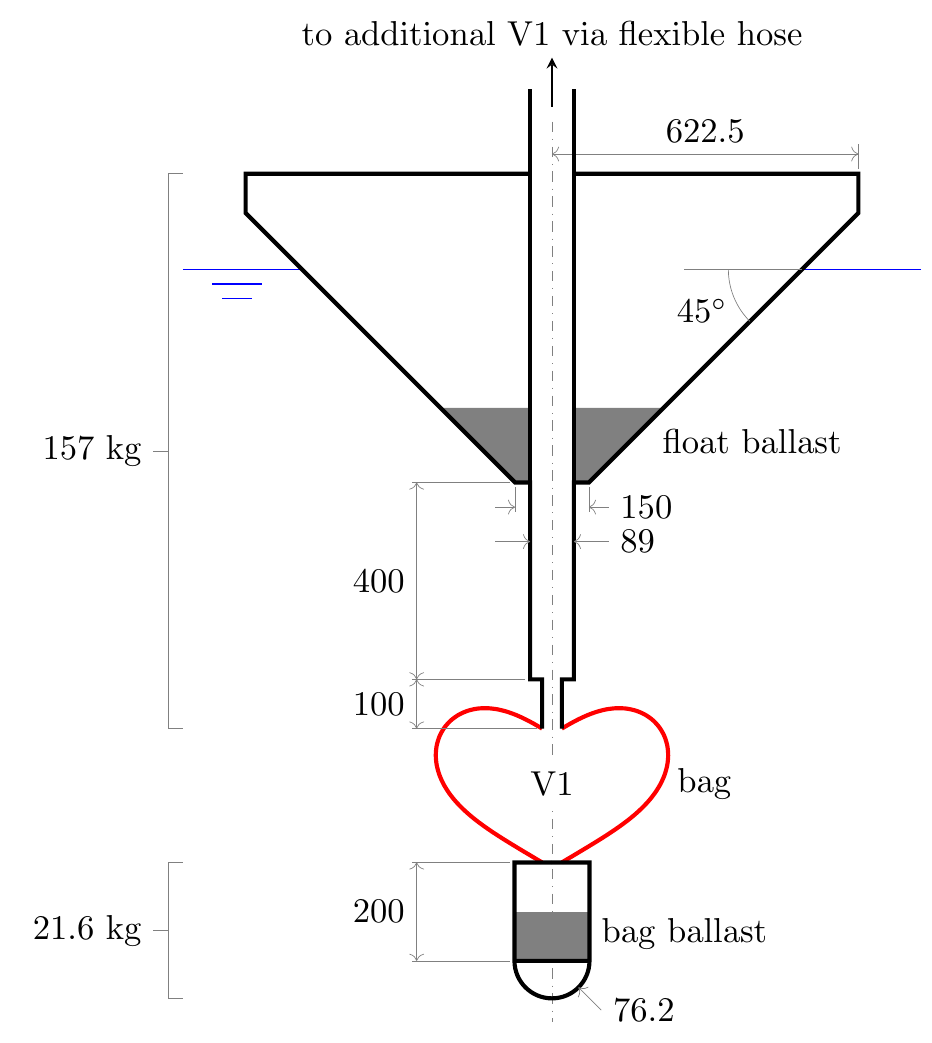}
\caption{Model dimensions (in mm).}
\label{testsetup}
\end{figure}

{The float was conical in shape and made of aluminium. 
As part of the float unit, a central PVC pipe ran through the float's vertical axis and was fixed to the float such that there was no relative motion between the two.
The pipe was open both at the top and bottom ends and functioned as a passage between the air volume contained within the bag and some additional air volume to be described in the following paragraph.
The bottom of the pipe was connected to the top of the bag via a brass skin fitting.} 
The float was ballasted {with lead shot} such that its mass, including the central pipe, was 157 kg.
{The bag was sealed at its bottom by an aluminium disc which was bolted to a cylinder made of aluminium and ballasted with lead shot.} 
The mass of this ballasted bottom cylinder was 21.6 kg.
The dimensions of the float and the bottom cylinder are shown in figure~\ref{testsetup}.

In order for the deformation of the model-scale bag to be similar to that of the full-scalle bag, the volumes of V1 and V2 must be scaled according to 
\begin{equation}
\frac{V_\mathrm{model}}{V_\mathrm{prototype}} =  \frac{1}{s^3}\frac{P_\mathrm{atm} s + P_\mathrm{prototype}}{P_\mathrm{atm} + P_\mathrm{prototype}} , \label{air_volume_scaling}
\end{equation}
where $s \geq 1$ is the scale factor, $P_\mathrm{atm}$ is the atmospheric pressure, and $P_\mathrm{prototype}$ is the mean internal pressure at full scale.
This means that $V_1$, i.e. the mean volume of V1, at model scale has to be greater than $1/s^3$ times the full-scale $V_1$.
{Therefore} during the tests, the volume of the bag was augmented with up to three air tanks {of} 220-litre capacity each{, to make up the total V1}. 
Another three tanks of the same capacity were used to make up V2.
{These tanks as well as the PTO were mounted on a gantry over the wave basin.
Thus in the model tests most of V1, the PTO, and all of V2 were external to the floating device.
A flexible hose connected the top of the central pipe to the tank assembly.
The force introduced by this flexible connection was minor compared to the static and dynamic force of the water acting on the device.}
The number of tanks in operation and thus the overall sizes of V1 and V2 could be varied by opening and closing of valves.
In the following, the size of V1 and V2 will be identified as V1 + V2 tanks = 1 + 1, 1 + 2, etc.

The PTO was in the form of an assembly of parallel capillary pipes between V1 and V2 {tanks} creating a linear resistance to the air flow. 
The assembly consisted of 17 tubes, each housing about 140 pipes of 1.6-mm internal diameter and  800-mm length.
{The same assembly was used previously for the model tests of another device}~\cite{Chaplin2012}. 
The resistance to the flow could be varied by opening and closing some of the tubes by means of valves.
The covered range of resistance was between 73 and 39 kPa m$^{-3}$ s, corresponding to 9 and 17 open tubes.

Linear guide rails similar to those used in~\cite{Beatty2015} restricted the {float} to move in heave only. 
A string potentiometer measured the vertical displacement of the float. 
Video cameras above and under water recorded the motion of the device from the side.
{Other instrumentation included a manometer to monitor the pressure in the system, pressure transducers to record the pressures in V1 and V2, and an array of wave gauges to record the free-surface elevation during the regular wave tests.}

\section{Results and discussions}

\subsection{Static trajectory}

\begin{figure}
\centering
\includegraphics[trim = 52mm 124mm 39mm 30mm, clip, scale=.9]{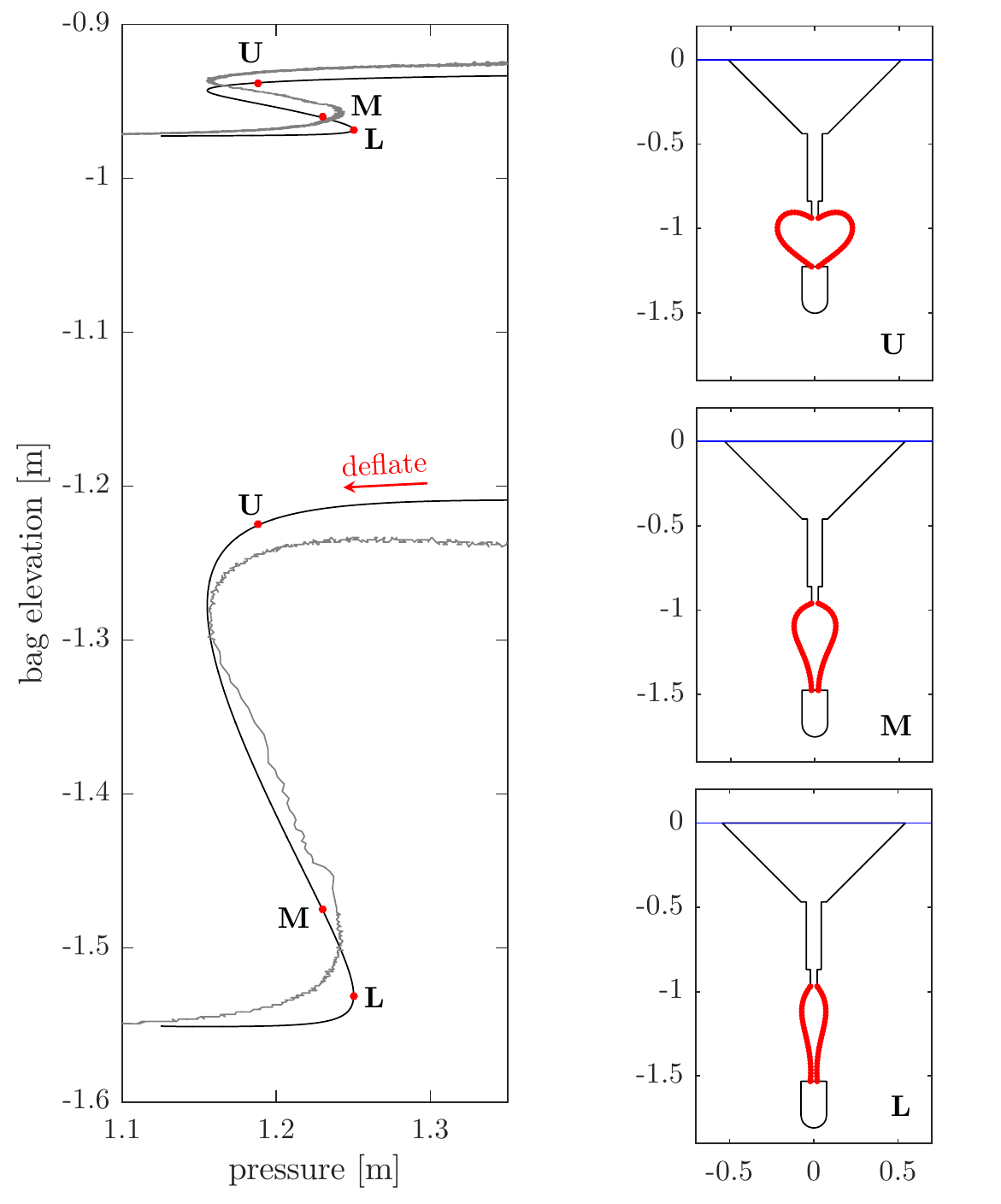}
\caption{Left: Measured (grey) and predicted (black) static trajectories of the top and bottom of the bag. 
Right: Calculated static geometries at marked points on the trajectories. Axes units are metres.}
\label{static_traj}
\end{figure}

We first examine the behaviour of the device in still water.
With the device floating freely, the bag is slowly deflated/inflated.
The measured variation of the top and bottom elevations of the bag with internal pressure is shown in figure~\ref{static_traj} along with the numerical predictions, which are the solutions of~\eqref{bagstatic} and any two of~\eqref{devicewholestatic},~\eqref{floatstatic}, or~\eqref{botcylstatic}.
The measured top trajectory was obtained from the string potentiometer connected to the float. 
The measured bottom trajectory was obtained from video images.
Calculated static geometries at marked points on the trajectories are also shown in figure~\ref{static_traj}.

The predicted trajectories agree favourably with the measurements. 
The discrepancies are believed to be the consequence of the relatively small number of tendons on the physical bag, resulting in the fabric sharing some of the tension which in the numerical model is carried exclusively by the tendons.
This is supported by Pagitz and Pellegrino~\cite{Pagitz2010}, who have shown, for a uniform external-internal pressure difference, that a bag with a finite number of tendons would have a slightly different profile than the profile of a bag with no hoop tension.
 
The static trajectories of the top and bottom elevations of the bag each takes the form of an S, which implies that for a range of bag pressures there are three possible bag shapes which have the same pressure.
The two {vertical-tangent} points of the S curve mark three different sections of the trajectory.
On the upper and lower sections, any pressure variation is accompanied with hardly any change in the device elevation. 
This is because on the upper section, the bag is tight, and releasing air from the bag decreases the pressure but hardly changes the bag's shape. 
Consequently, there is not much change in the device's draught in order to maintain buoyancy.  
Likewise, on the lower section, the tendons are close to vertical and deflating the bag decreases the pressure but hardly changes the bag's volume, and so the device's draught stays about the same.
Between these two extremes, i.e. on the middle section,  varying the amount of air in the bag has the greatest effect on the bag shape and hence there is greater variation in the device draught in order to maintain buoyancy. 
Contrary to the upper and lower sections, however, on the middle section deflating the bag does not decrease the pressure but instead increases it. 
The reason is that as the device goes down to provide the required buoyancy, the external hydrostatic pressure increases, and this increases the pressure in the bag.

For a  ballasted floating bag (without a rigid float)~\cite{Kurniawan2017}, the static trajectory takes the form of a C, with a shape similar to the combined upper and middle sections of the S trajectory of the present device. 
The existence of the lower section in the present trajectory is due to the extra buoyancy provided by the float, preventing the device from sinking even when the bag has been completely deflated.

\subsection{Static stability}

For a given waterplane radius, the upward force on the float due to the bag must not exceed a certain maximum in order to keep the device from capsizing. 
A simplified expression for this maximum may be derived for our device by assuming the float to be a simple $45^\circ$ cone with waterplane radius $R_{\mathrm w}$, and neglecting the central pipe. 
The centre of buoyancy of the float is thus at $R_{\mathrm w}/3$ below the waterline. 
Denoting $d_\mathrm{G}$ to be the distance from the bottom of the float (i.e. the apex of the cone) to  the float's centre of gravity,  and $d_\mathrm{T}$ to be the distance from the top of the bag to the bottom of the float, and assuming that $0 \leq d_\mathrm{G} \leq R_{\mathrm w}$, we can show that for stability the vertical component of the sum of tension at the top of the bag has to be
\begin{equation}
T_\mathrm{v} \equiv T \sin A_1 < \frac{\rho g \pi R_{\mathrm w}^4 }{d_\mathrm{G} + d_\mathrm{T}} \left( \frac{17}{36} - \frac{d_\mathrm{G}}{3R_{\mathrm w}} \right). \label{eq_stability}
\end{equation}
Equation~\eqref{floatstatic}, on the other hand, gives the required vertical component of the tendon tension in order to satisfy force equilibrium on the float.

\begin{figure}
\centering
\includegraphics[trim = 52mm 221mm 50mm 20mm, clip, scale=.9]{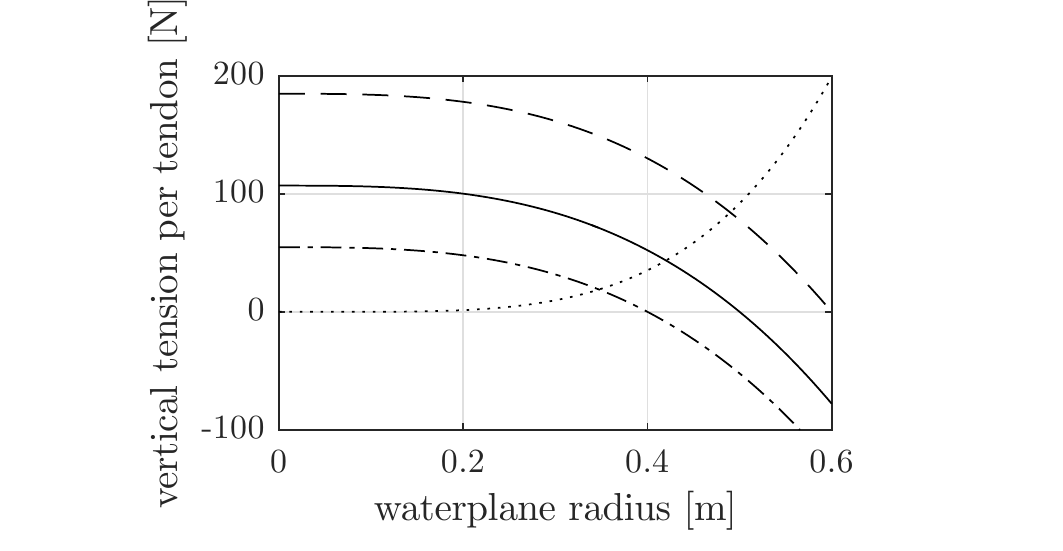}
\caption{Vertical component of tendon tension for equilibrium assuming $R_\mathrm{top} = 0$ and $M_\mathrm{float} = \rho \pi 0.5^3 / 3$ kg (solid), $M_\mathrm{float} = \rho \pi 0.6^3 / 3$ kg (dashed), and $M_\mathrm{float} = \rho \pi 0.4^3 / 3$ kg (dash-dotted). 
Maximum allowable vertical force for stability is shown as the dotted line. 
The centre of gravity of the float is assumed to be 0.15 m above the cone's lowest point ($d_\mathrm{G} = 0.15$ m). 
The bag top is assumed to be 0.5 m below the cone ($d_\mathrm{T} = 0.5$ m).}
\label{static_stability}
\end{figure}

The two equations are plotted in figure~\ref{static_stability}.  
Equation~\eqref{eq_stability} is shown as the ascending dotted line, while equation~\eqref{floatstatic}, with $R_\mathrm{top} = 0$ and for different values of $M_\mathrm{float}$, is shown as the descending lines.
The intersection of the ascending line and any of the descending lines gives the maximum allowable vertical force on the float due to the bag, and the corresponding minimum waterplane radius, for a given float mass. 
At any point right of the intersection of the two lines, the device is stable since the required vertical force to satisfy equilibrium is less than the allowable maximum for stability.
\begin{figure}
\centering
\includegraphics[trim = 0mm 3mm 0mm 1mm, clip, scale=.9]{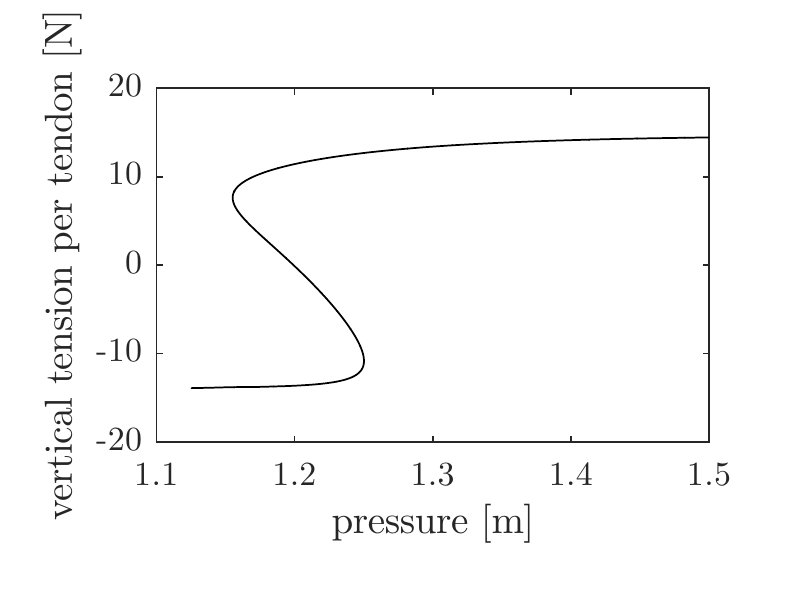}
\caption{Trajectory of the vertical component of tendon tension at the top of the bag, for the model in figure~\ref{testsetup}. 
The vertical force is positive if the bag pushes the float up and negative if the bag pulls the float down.}
\label{verticaltension}
\end{figure}
Consider, for example, the solid line, which corresponds to $M_\mathrm{float} = \rho \pi 0.5^3 / 3$ kg.
As the bag is inflated from a deflated state, we move along the solid line from right to left.
At waterplane radius $R_{\mathrm w}$ of 0.5 m, the vertical force on the float exerted by the bag is zero and the tendons leave the top of the bag in the horizontal direction ($A_1 = 0$). 
To the right of this point ($R_{\mathrm w} > 0.5$ m), the bag pulls the float down, while to the left ($R_{\mathrm w} < 0.5$ m), the bag pushes the float up. 
The bag can be inflated until the float's waterplane radius reaches 0.42 m, but 
further inflating the bag past this point will cause the device to capsize. 

As seen from~\eqref{eq_stability}, the onset of instability may be delayed by lowering the centre of gravity of the float or reducing the distance between the bottom of the float and the top of the bag. 
Whether the vertical force on the bag will ever exceed the maximum allowable force to maintain stability of the device depends on the buoyancy of the bag relative to the float's weight. 
For the model shown in figure~\ref{testsetup}, the vertical upward force per tendon exerted on the float is found to be less than 20 N (figure~\ref{verticaltension}).  
The device is therefore always stable. 
With a lower float mass, as seen from figure~\ref{static_stability}, the maximum allowable force would be lower. 


\subsection{Heave natural periods}

The heave natural period of the device, when the bag is fully deflated, was measured by forcing the device to oscillate in otherwise still water.
The measured natural period, 1.19 s, agrees closely with the numerically predicted 
natural period, 1.17 s.  

\begin{figure}
\centering
\includegraphics[trim = 35mm 227mm 35mm 21mm, clip, scale=.9]{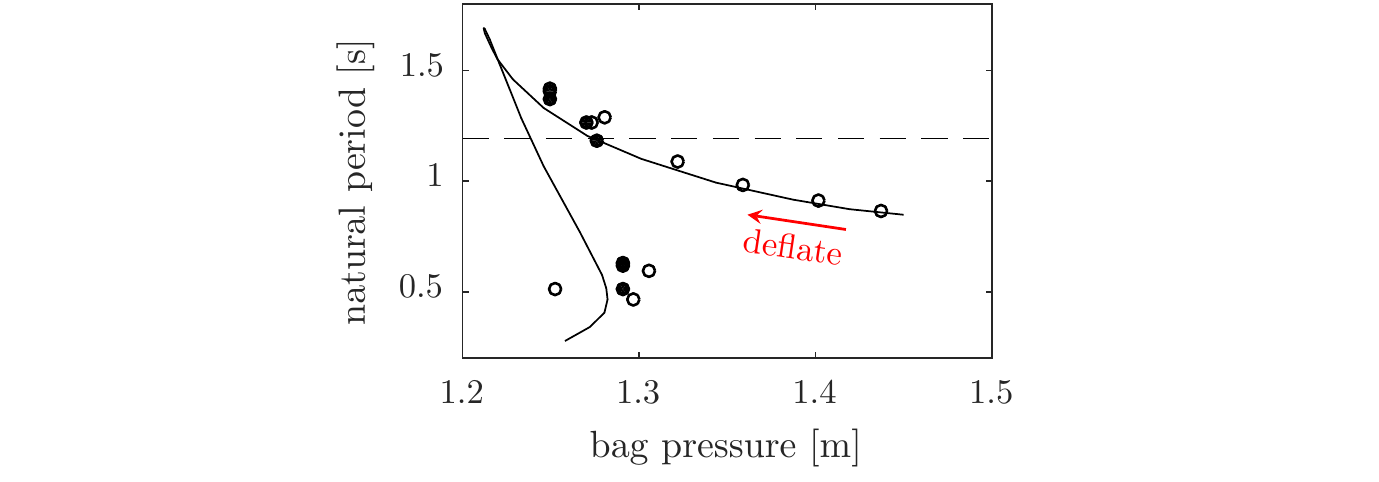}\\
\caption{{Predicted (solid line) and measured (circles)} heave natural period of the bottom cylinder with the float fixed in position such that the top of the bag is 1 m below the mean water line. 
V1 + V2 tanks = 1 + 3 and 9 PTO tubes are open.
Open and filled circles correspond to two different runs of the test. 
{The natural period is shown to vary depending on the amount of air in the bag.}
{As a comparison,} the measured heave natural period of the device with the bag fully deflated is shown as the horizontal dashed line.}
\label{P-T0bal}
\end{figure}

The bag essentially acts as a spring, whose stiffness can be adjusted by varying the amount of air in the bag.
Such a spring may have applications beyond what is suggested in this paper. 
Figure~\ref{P-T0bal} shows a typical variation of peak period of the heave displacement of the bottom cylinder with bag pressure, as the bag was slowly deflated. 
In this case, the float was fixed in position such that the top of the bag was always 1 m below the mean free surface.
The bottom cylinder was forced to oscillate via a line connected to the base of the cylinder, in otherwise still water.
The number of V1 + V2 tanks was 1 + 3, and 9 PTO tubes were open.
The frequency-domain model predicts the measured heave natural periods of the bottom cylinder reasonably well.
Following the trajectory from the right, we see that with decreasing amount of air in the bag, the natural period of the bottom cylinder first increases until it reaches a maximum, and then decreases with further bag deflation.
Depending on the amount of air in the bag, the heave natural period of the bottom cylinder varies over quite a wide range.


\subsection{Response in regular incident waves}

\begin{figure}
\centering
\includegraphics[trim = 5mm 2mm 2mm 3mm, clip,  width=\textwidth]{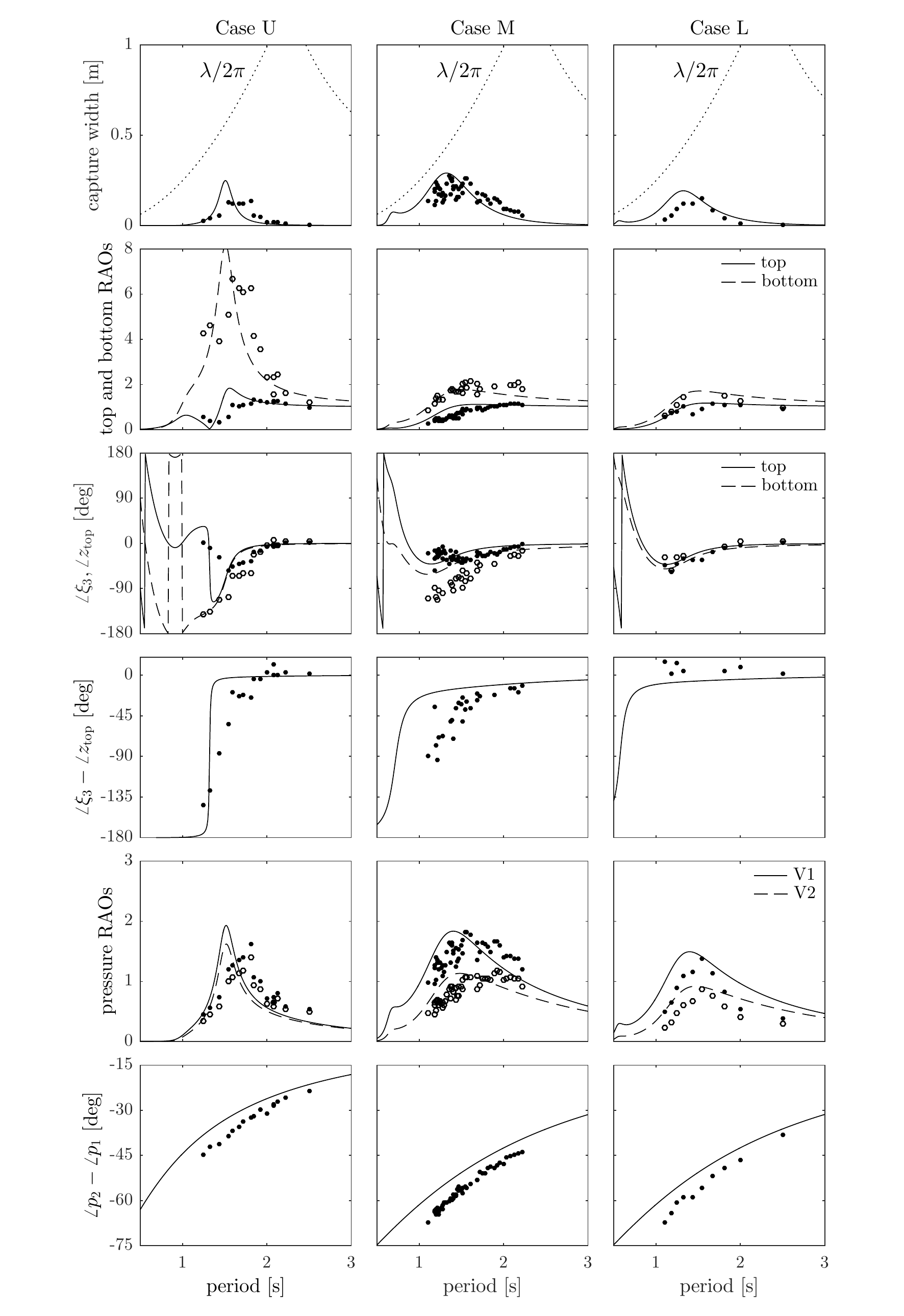}
\caption{{Predicted (lines) and measured (circles)} response of the device in regular incident waves, for cases specified in table~\ref{cases}. 
The {top and bottom response amplitude operators (RAOs) are defined as the displacement amplitudes of the top and bottom of the bag, normalised by the incident wave amplitude $A$.}
{The pressure RAOs are the V1 and V2 pressure amplitudes in metres of water normalised by $A$}. 
{In each capture width plot, the ascending dotted line is the $\lambda/2\pi$ upper bound, where $\lambda$ is the incident wavelength, while the descending dotted line is the Budal upper bound assuming a design volume stroke of $2AS_\mathrm{w}$, where $S_\mathrm{w}$ is the device waterplane area.}}
\label{incwaveresults}
\end{figure}

The response of the device in regular incident waves, for three different mean conditions as listed in table~\ref{cases}, is shown in figure~\ref{incwaveresults}.
These mean conditions correspond to the marked points on the static trajectory shown earlier in figure~\ref{static_traj}.
The incident wave amplitude used in the tests was 3 cm for all cases.

\begin{table}[]
\centering
\caption{Description of the test cases.}
\label{cases}
\begin{tabular}{cccccc}
\toprule
Test & V1 + V2 & PTO & Mean position & mean & mean bag \\
case & tanks & tubes & on S trajectory & pressure [m] & volume [L] \\
\midrule 
U 	& 1 + 3      & 17       	& upper	& 1.1881 & 30.754	\\ 
M 	& 1 + 3      & 9       	& middle	& 1.2301 & 11.165	\\ 
L    	& 3 + 3	& 9		& lower 	& 1.2502 & 3.986	\\ 
\bottomrule
\end{tabular}
\end{table}

Different responses are obtained for different mean conditions. 
Slight variation in the initial condition from one run to another contributes to the scatter of the measurements, but fairly good agreement is obtained between the measurements and the numerical predictions.
Sensitivity analysis with slightly different mean pressures as input to the numerical model does not improve the agreement, suggesting that the discrepancies, most notably the mismatch in peak periods, might be a consequence of geometrical differences between the physical model and the numerical model, which {are manifested} also in the differences between the predicted and measured static trajectories observed earlier.
Nevertheless, the numerical model captures reasonably well the effect of varying parameters on the response of the device. 

In particular, there is a good agreement between the measured and predicted phase difference of the pressure amplitudes in V1 and V2, for all cases. 
The numerical model confirms the linearity of the PTO damping in the physical model, although 
the measured phase differences are slightly greater than the predictions. 
This is contributed by additional losses in the PTO, which are not accounted for in the numerical model.
As expected, increasing the number of open PTO tubes from 9 to 17 decreases the phase difference between pressures in the two volumes as well as  the amplitude difference between the two pressures.

Despite the measured response peak periods being higher than the predictions, the amplitudes of the top and bottom displacements of the bag are well-predicted by the numerical model. 
The corresponding phase variations are also predicted reasonably well. 
The observed phase shift of the bottom of the bag relative to the top happens at decreasing period from case U to L.
This period coincides with the heave natural period of the bottom cylinder.
At this period the float hardly moves, which is most clearly seen for case U at 1.32 s.  

\begin{figure}
\centering
\includegraphics[trim = 18mm 216mm 14mm 0mm, clip,  width=\textwidth]{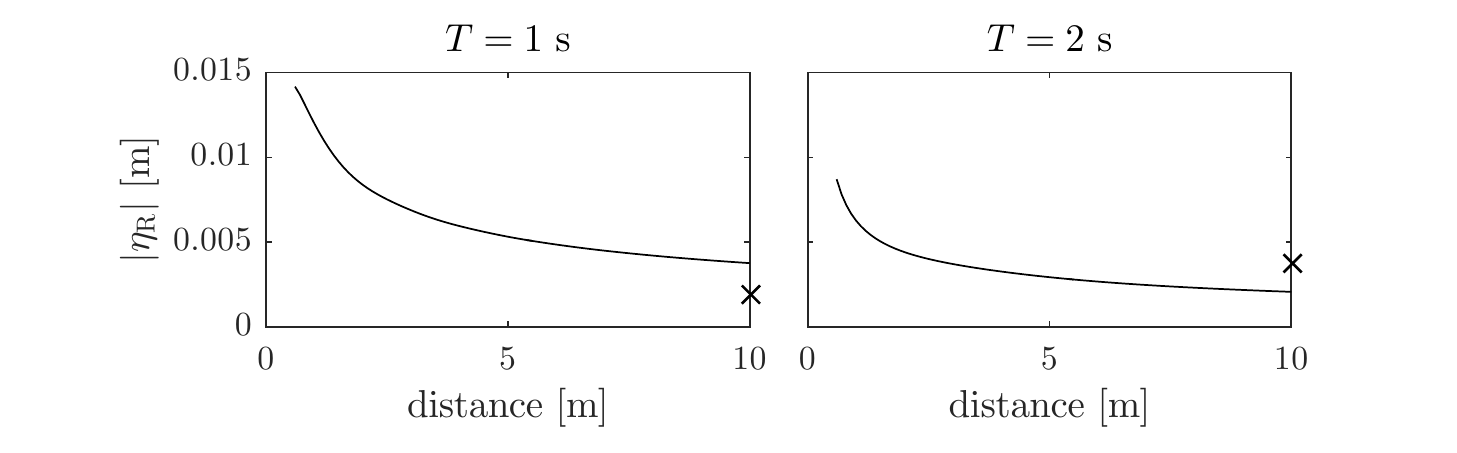}
\caption{{Calculated radiated wave amplitudes at $T = 1$ s and $T = 2$ s, for case U in table~\ref{cases}. 
The optimum radiated wave amplitudes at a distance of 10 m from the device's vertical axis are indicated with crosses.}}
\label{caseUetaR}
\end{figure}

Operating with a mean condition on the upper section of the S trajectory (case U) is found to result in a relatively poor power performance. 
This is rather surprising, given that the relative displacement between the bottom cylinder and the float is much greater in this case compared to that found in cases M and L.
{We observe that,
firstly,} it is possible for the bag to change its shape without much change in its volume. 
{Power absorption however requires a certain volume flow through the PTO.
Secondly, to absorb energy from the waves a device must radiate waves that interfere destructively with the incident and scattered waves (see, e.g.~\cite{Falnes2015}).} 
The low absorbed power either side of the peak period implies that {either the  amplitude or the phase, or both, of the radiated waves is not optimal}.  
{A closer inspection reveals that at short periods
the waves radiated by the device for case U are too high, while at long periods they are too low (see figure~\ref{caseUetaR}).
The phases of these radiated waves are probably also suboptimal.}

{To obtain the optimum radiated wave amplitudes in figure~\ref{caseUetaR}, we have made use of the fact that for a device to absorb the maximum amount of power possible from the waves, it is necessary for it to radiate waves having exactly the same amount of power. 
The power of the radiated waves for this axisymmetric device can be approximated as
\begin{equation}
\mathcal{P}_\mathrm{R} \approx 2 \pi r  \frac{1}{2} \rho g    |\eta_\mathrm{R}(r)|^2  v_\mathrm{g}  ,
\end{equation}
where $|\eta_\mathrm{R}(r)|$ is the radiated wave amplitude at a sufficiently large distance $r$ from the device and $v_\mathrm{g}$ is the wave group velocity. 
For an axisymmetric device, the maximum power that it can absorb from incident regular waves of amplitude $A$ is given theoretically as 
\begin{equation}
\mathcal{P}_\mathrm{max} = \frac{\lambda}{2\pi}J  = \frac{\lambda}{2\pi} \frac{1}{2} \rho g A^2 v_\mathrm{g} , 
\end{equation}
where $\lambda$ is the incident wavelength and $J = \frac{1}{2} \rho g A^2 v_\mathrm{g}$ is the transported power per unit width of the incident wave front~\cite{Budal1975,Newman1976,Evans1976}.
Here, $\lambda /2\pi$ is the well-known upper bound of the capture width, which has been plotted in figures~\ref{incwaveresults},~\ref{V1V2study},~\ref{extracases}, and~\ref{radiation_case}.
The optimum radiated wave amplitude at a sufficiently large distance $r$ from the device can thus be obtained by equating $\mathcal{P}_\mathrm{R\,opt} = \mathcal{P}_\mathrm{max}$, which gives
\begin{equation}
2 \pi r |\eta_\mathrm{R\,opt}|^2 \approx \frac{ \lambda}{2\pi} A^2  .
\end{equation}
}

The device is found to absorb more power when operating with a mean condition on the middle section of the trajectory (case M). 
It will be shown in the next section that improved power capture can be obtained {in this case} by increasing the sizes of V1 and V2.

{For all cases,} the float and the bottom cylinder are found to move in phase at long periods and in antiphase at short periods.
Such behaviour is characteristic for a two-degree-of-freedom (2-DOF) system. 
Indeed, we can think of the device as a system of two rigid bodies connected by a spring-damper element in the form of an air bag.
However, while a simple 2-DOF system is able to explain this phase shift between the float and the bottom cylinder, sufficiently accurate predictions of the power absorbed by the device seem to be possible only with a model which takes into account the deformations of the bag and its interactions with the two rigid bodies, as used in this paper. 
Such interactions are more subtle than what can possibly be modelled by a simple spring-damper system.

\subsection{Parametric study} \label{parametric_study}

Having validated the numerical model to some extent, we will now use it to 
better understand the behaviour of the device  for a wider range of conditions and parameters.

We start with the cases specified in table~\ref{cases}, but with different values of mean volumes $V_1$ and $V_2$ and PTO damping to see how they affect the power absorption.

\begin{figure}
\centering
\includegraphics[trim = 16mm 204mm 14mm 3mm, clip,  width=\textwidth]{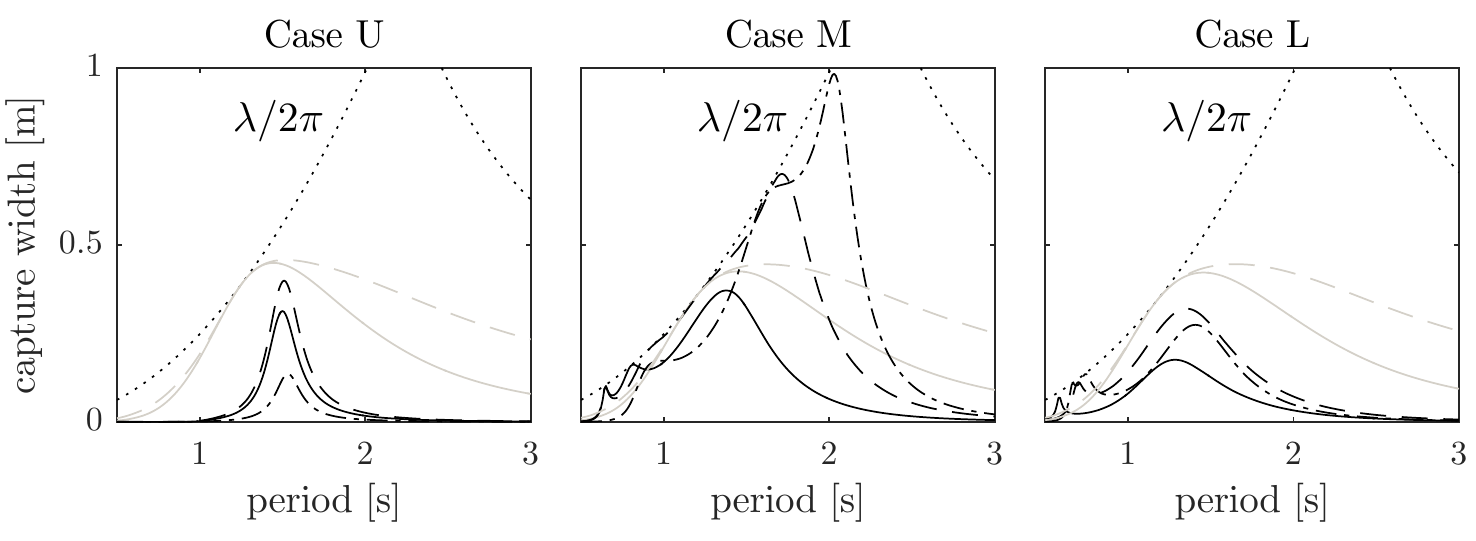}
\caption{Calculated capture widths for cases specified in table~\ref{cases} but with V1 + V2 tanks = 1 + 3 (black solid), V1 + V2 tanks = 1 + 6 (black dashed), V1 + V2 tanks = 6 + 3 (black dash-dotted), and with PTO damping optimised at every period. For each case, the grey lines are the capture widths of a rigid body of the same mean geometry absorbing power through heave relative to a fixed reference, with PTO damping equal to the heave radiation damping at resonance (grey solid) and with PTO damping optimised at every period (grey dashed).
The ascending dotted lines are the $\lambda/2\pi$ upper bound, {while the descending dotted lines are the Budal upper bounds assuming a design volume stroke of $2AS_\mathrm{w}$ (see caption of figure~\ref{incwaveresults} for explanation of notations)}.}
\label{V1V2study}
\end{figure}

For case U, we find that increasing $V_2$ and decreasing $V_1$ with the same PTO damping improve power absorption, but $V_2$ must be at least 3 times larger than that specified in table~\ref{cases} for the peak absorbed power to reach 80\% of the theoretical maximum. 
Furthermore, no combinations of $V_1$, $V_2$, and PTO damping are found to improve the absorption bandwidth (see figure~\ref{V1V2study}), suggesting that the narrow bandwidth is the consequence of the shape of the bag associated with this particular mean condition. 

The performance is slightly better for case L, but $V_1$ and $V_2$ must be very, and perhaps impractically, large for the absorbed power to be comparable to that of an equal-sized rigid body absorbing power through heave relative to a fixed reference.
 
Out of  the three cases, the highest power absorption is obtained for case M. 
The capture width is found to increase with larger $V_1$ and $V_2$.
Although the bandwidth is narrower than that of an equal-sized heaving rigid body, the capture width can attain values higher than 40\% of the waterplane diameter, which is the achievable maximum for the rigid body. 

It must be borne in mind, however, that all the results presented here have been obtained based on linear theory. 
In order to realise the capture width shown by the dash-dotted line for case M in figure~\ref{V1V2study}, for example, the relative displacement between the float and the bottom cylinder needs to be up to 9 times the incident wave amplitude. 
Since the distance between the top and bottom of the bag is limited by the tendon length, this result is valid only for very small incident wave amplitudes.

\begin{figure}
\centering
\includegraphics[trim = 52mm 124mm 39mm 30mm, clip, scale=.9]{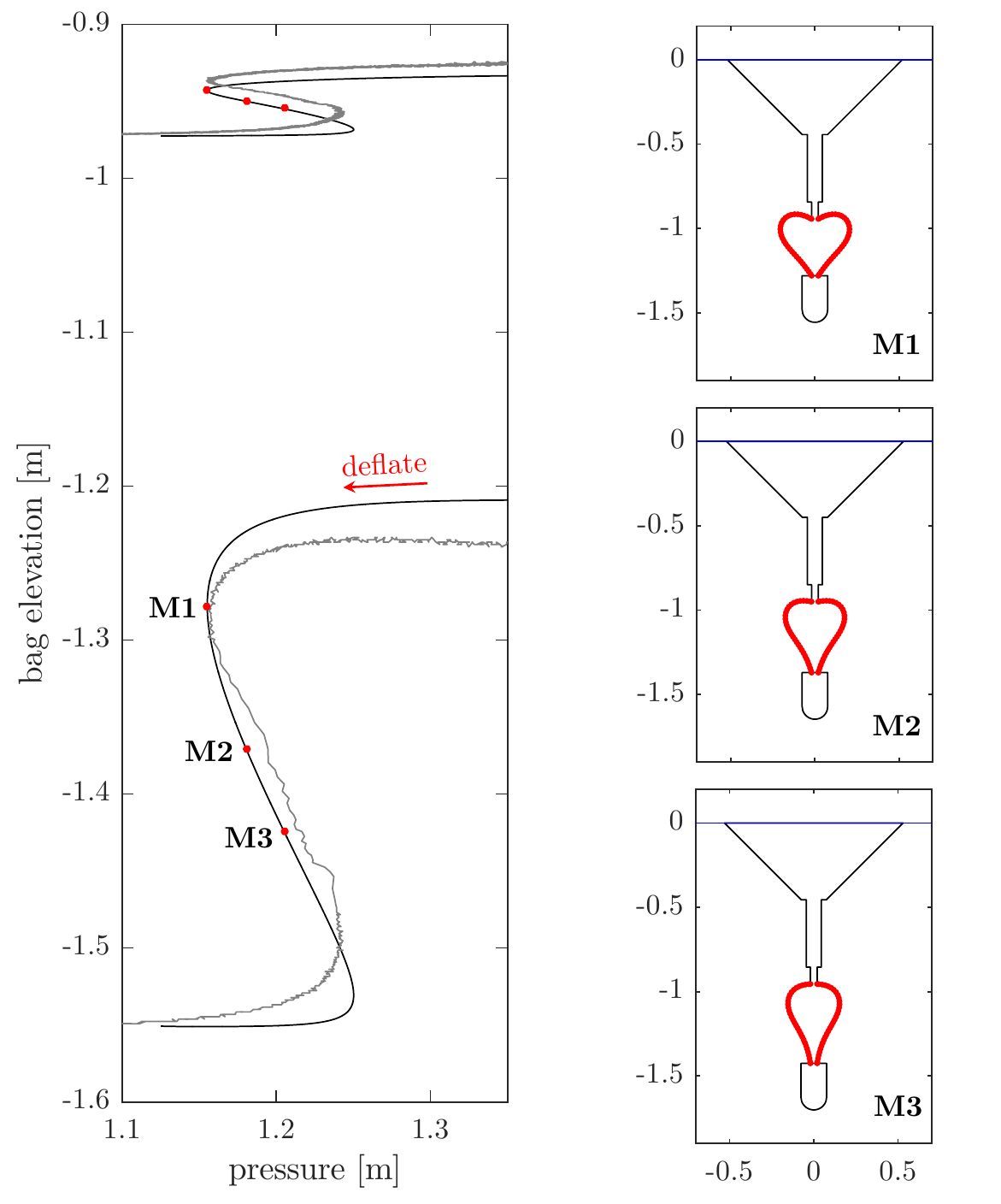}
\caption{Additional cases as specified in table~\ref{casesm}, their positions on the trajectories, and the corresponding mean geometries. Axes units are metres.
Refer to figure~\ref{static_traj} for explanation of line types in the trajectory plot.}
\label{static_trajm}
\end{figure}

\begin{table}[]
\centering
\caption{Description of the extra cases with mean conditions on the middle section of the S trajectory.}
\label{casesm}
\begin{tabular}{ccc}
\toprule
\multirow{2}{*}{Case} & mean & mean bag \\
 & pressure [m] & volume [L] \\
\midrule 
M1	& 1.1552 & 26.522	\\ 
M2 	& 1.1802 & 20.625	\\
M3 	& 1.2051 & 16.250	\\ 
\bottomrule
\end{tabular}
\end{table}

\begin{figure}
\centering
\includegraphics[trim = 3mm 3mm 1mm 1mm, clip,  width=.4\textwidth]{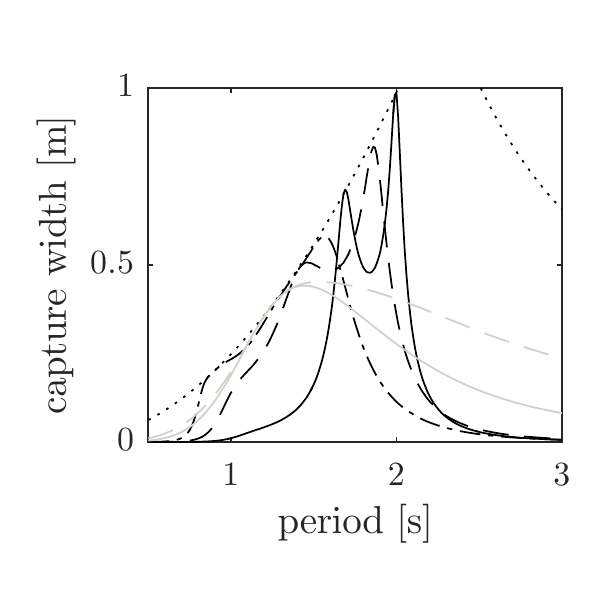}
\caption{Calculated capture widths for cases specified in table~\ref{casesm} with V1 + V2 tanks = 1 + 3 and with PTO damping optimised at every period: 
case M1 (solid), case M2 (dashed), case M3 (dash-dotted). 
Also shown are the capture widths of a rigid body absorbing power through heave relative to a fixed reference, with PTO damping equal to the heave radiation damping at resonance (grey solid) and with PTO damping optimised at every period (grey dashed).
The rigid body has the same mean geometry as that of the device in case M1. 
The capture widths of rigid bodies with the same mean geometries as those of the device in cases M2 and M3 are similar.
The ascending dotted line is the $\lambda/2\pi$ upper bound{,
while the descending dotted line is the Budal upper bound for case M2, assuming a design volume stroke of $2AS_\mathrm{w}$
 (see caption of figure~\ref{incwaveresults} for explanation of notations).
 The Budal upper bounds for cases M1 and M3 are similar.}
}
\label{extracases}
\end{figure}

To see how the response of the device varies for different mean conditions on the middle section of the static trajectory, we consider three additional cases with mean pressures as listed in table~\ref{casesm} and whose locations in the trajectory  are shown in figure~\ref{static_trajm}. 
Figure~\ref{extracases} shows the calculated capture widths of the device for these mean conditions.

For cases on the middle section of the trajectory,  the capture widths generally have two peaks when constant PTO dampings are used.
As the PTO damping is reduced, the peaks shift towards longer periods. 
This is because in the limit of no flow resistance between V1 and V2, the two volumes in effect become one single volume. 
Thus the heave restoring stiffness of the bottom cylinder is reduced with decreasing PTO damping, resulting in longer resonance period.
When the envelope of all capture widths obtained using constant PTO damping is taken, which is equivalent to the capture width with PTO damping optimised at every period,
it generally has three peaks. 
Increasing the total available air volumes $V_1 + V_2$ increases the highest peak period, whereas the lowest peak period is more or less governed by $V_1$  (cf. figure~\ref{V1V2study}).

Starting with different mean pressures, as seen from figures~\ref{V1V2study} and~\ref{extracases}, results in distinct responses. 
In agreement with figure~\ref{P-T0bal}, figure~\ref{extracases} shows that going down the middle section of the static trajectory shifts the response curve to lower periods. 
It appears that the best power performance is obtained when the bag exerts no mean vertical force on the float, i.e. between cases M2 and M3. 



\begin{figure}
\centering
\includegraphics[trim = 3mm 3mm 1mm 1mm, clip,  width=.4\textwidth]{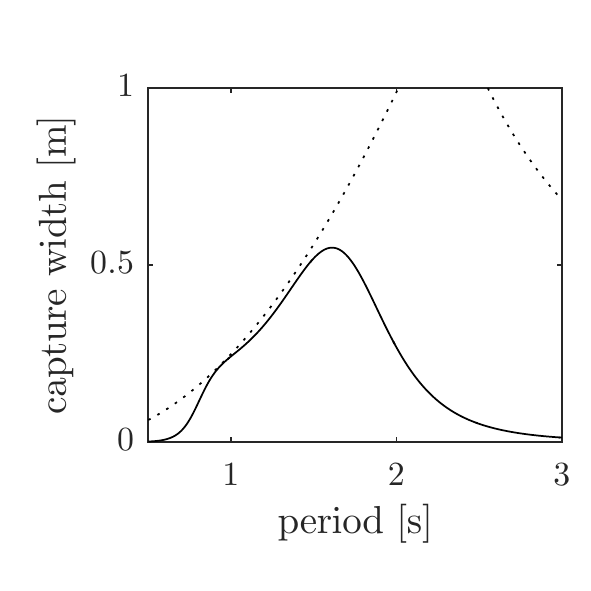}
\caption{{Calculated capture width for case M in table~\ref{cases}, with V1 + V2 tanks = 1 + 6 and 45 open PTO tubes. 
The ascending dotted line is the $\lambda/2\pi$ {upper bound}, while the descending dotted line is the Budal {upper bound}  assuming a design volume stroke of $2AS_\mathrm{w}$ (see caption of figure~\ref{incwaveresults} for explanation of notations).}}
\label{radiation_case}
\end{figure}

\begin{figure}
\centering
\includegraphics[trim = 18mm 206mm 14mm 3mm, clip,  width=\textwidth]{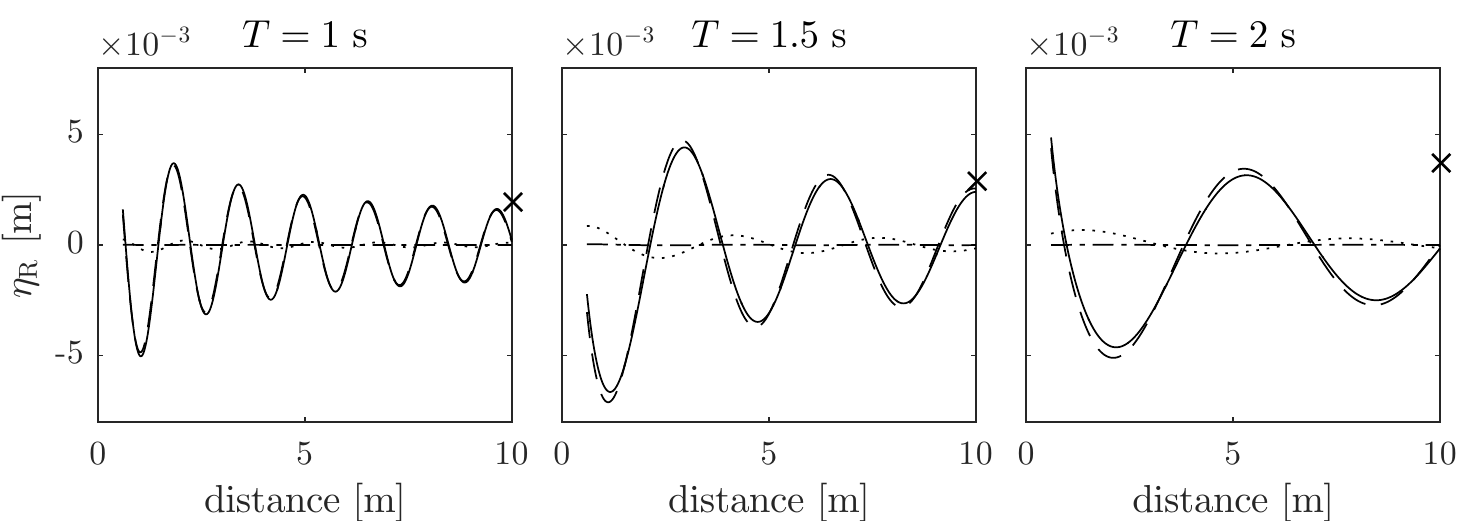}
\caption{{Calculated total radiated wave (solid) and radiated wave components due to the float (dashed), bag (dotted), and bottom cylinder (dash-dotted), for the case in figure~\ref{radiation_case}, at three different wave periods and  time $t = 0$ second.
The optimum total radiated wave amplitudes at a distance of 10 m from the device's vertical axis are indicated with crosses.}}
\label{radiation_components}
\end{figure}

{Before concluding, we shall make two remarks. 
First, alongside  the $\lambda/2\pi$ upper bound, we have plotted   the Budal upper bound~\cite{BudalFalnes1980} in the capture width plots of figures~\ref{incwaveresults},~\ref{V1V2study},~\ref{extracases}, and~\ref{radiation_case}.   
These Budal {upper bounds} are plotted with the sole purpose of showing that the power a device can absorb is limited by its dimensions, and are not meant to give any indications about the economics of the device.
They have been derived assuming that the device is a heaving body with a design volume stroke, i.e. twice the displaced volume amplitude, of $2AS_\mathrm{w}$, where $A$ is the incident wave amplitude and $S_\mathrm{w}$ is the device waterplane area.
This choice of the design volume stroke is reasonable given the float's response amplitude operator (RAO) shown in figure~\ref{incwaveresults} and the fact that the displaced volume amplitude of the bag tends to subtract that of the float.
Thus, the Budal {upper bound}  in terms of capture width in this case becomes
\begin{equation}
d_\mathrm{a} < \frac{2\pi S_\mathrm{w}}{T v_\mathrm{g}} . 
\end{equation}
Notice that by assuming the design volume stroke to be proportional to the incident wave amplitude, we have arrived at an expression independent of the incident wave amplitude.
In reality, the design volume stroke would be a constant, and thus the incident wave amplitude would enter into the expression. 
In this case, the Budal {upper bounds} in figures~\ref{incwaveresults},~\ref{V1V2study},~\ref{extracases}, and~\ref{radiation_case} may be understood as the  {upper bounds} applicable for a design volume stroke of $0.06 S_\mathrm{w}$ m$^\text{3}$ and $A = 0.03$ m.
{(Note that to reach some of the optimised results shown in figures~\ref{V1V2study} and~\ref{extracases}, the actual volume stroke may need to exceed this particular design stroke.) }
As seen from figures~\ref{V1V2study} and~\ref{extracases}, an equal-sized heaving rigid body reacting against a fixed reference can still absorb some amount of power at long periods, whereas the absorbed power of the present device, because it is self-reacting, drops quite rapidly at long periods. 

Secondly, we shall comment on the relative contributions of the parts of the device---the float, the bag, and the bottom cylinder---in generating the total radiated waves.
Typical contributions are shown in figure~\ref{radiation_components}, corresponding to the case described in figure~\ref{radiation_case}. 
We see that the contribution of the bottom cylinder is negligible, while the bag's contribution is small compared to the float's. 
It is worth noting, however, that the waves radiated by the bag (and the bottom cylinder) tend to reduce those radiated by the float, and this is another reason why the bandwidth of the device is narrower than that of an equal-sized rigid body. 
}

\section{Conclusion}

The static and dynamic behaviour of a novel wave energy device featuring a submerged flexible air bag connected at the top to a ballasted float and at the bottom to a ballasted cylinder, have been investigated numerically as well as by physical experiments. 
The device is one variation of a newly emerging class of devices which have in common a flexible bag in the form of a fabric encased within an array of longitudinal tendons.

In the numerical models, the assumption is that the tendons carry all the tension and that there is no tension in the fabric. 
The profile of the bag is obtained by discretising the tendon into small contiguous elements, able to move relative to each other as long as the distance between centres of any two neighbouring elements does not change. 
Full linear hydrodynamic interactions between the float, the bottom cylinder, and the deformable bag are accounted for.
{Comparisons with the physical measurements show favourable agreement.}

In still water, static equilibrium is achieved when the weight of the device is equal to its buoyancy. 
Depending on the amount of air in the bag, the bag can take different shapes and exert either an upward or downward force on the float. 
Plotting the static elevations of the top or bottom of the bag against the bag pressure results in an S-shaped trajectory, implying that there can be three different equilibrium device geometries for the same bag pressure.  
  
Dynamically, the device exhibits different behaviours depending on where the mean pressure-elevation condition is on the static trajectory. 
The highest power absorptions are obtained with mean conditions on the middle section {of the trajectory}, around the point where the bag exerts no mean vertical force on the float.
The power absorption can peak at a  period longer than the peak period of a heaving equal-sized rigid body reacting against a fixed reference, albeit having a narrower bandwidth.

Further study may include optimisations of device parameters such as the submergence of the bag from the mean water level; the relative sizes of the float, bag, and bottom cylinder; and the mass ratio between the float and the bottom cylinder.


\dataccess{The datasets supporting this article have been uploaded as part of the supplementary material.}

\aucontribute{AK carried out the numerical study, participated in the physical experiments, and drafted the manuscript; JRC devised the numerical approach for solving the problem, participated in the physical experiments, and helped draft the manuscript. MRH designed and participated in the physical experiments; DMG coordinated the study, participated in the physical experiments, and helped draft the manuscript; FJMF conceived the idea of the device and participated in the experiments.} 

\competing{We have no competing interests.}

\funding{This work is supported by the EPSRC SuperGen Marine Energy Research Consortium [EP/K012177/1]. }

\ack{We are grateful to Malcolm Cox of Griffon Hoverwork Ltd for supplying the model-scale bags and for useful discussions, and to the referees for their constructive suggestions.}

\bibliographystyle{vancouver}
\bibliography{wavenergy}{}

\begin{thebibliography}{10}

\bibitem{French1977}
French MJ.
\newblock Hydrodynamic basis of wave-energy converters of channel form.
\newblock Journal of Mechanical Engineering Science. 1977;19(2):90--92.

\bibitem{French1979}
French MJ.
\newblock The search for low cost wave energy and the flexible bag device.
\newblock In: Proc. 1st Symp. Wave Energy Utilization. Gothenburg, Sweden;
  1979. p. 364--377.

\bibitem{Kurniawan2016}
Kurniawan A, Greaves D, Hann M, Chaplin JR, Farley F.
\newblock Wave energy absorption by a floating air-filled bag.
\newblock In: Proc. 31st Int. Workshop on Water Waves and Floating Bodies.
  Plymouth, US; 2016. .

\bibitem{Kurniawan2017}
Kurniawan A, Chaplin JR, Greaves DM, Hann M.
\newblock Wave energy absorption by a floating air bag.
\newblock Journal of Fluid Mechanics. 2017;812:294--320.

\bibitem{Kurniawan2016b}
Kurniawan A, Greaves D.
\newblock Wave power absorption by a submerged balloon fixed to the sea bed.
\newblock IET Renewable Power Generation. 2016;10(10):1461--1467.

\bibitem{Farley2016patent}
Farley FJM. Wave power converter; 2016.
\newblock {Patent application GB2532074, filing date: 9 November 2014,
  publication date: 11 May 2016}.

\bibitem{SarmentoFalcao1985}
Sarmento AJNA, Falc\~ao AFdO.
\newblock Wave generation by an oscillating surface-pressure and its
  application in wave-energy extraction.
\newblock Journal of Fluid Mechanics. 1985;150:467--485.

\bibitem{MalmoReitan1985}
Malmo O, Reitan A.
\newblock Wave-power absorption by an oscillating water column in a channel.
\newblock Journal of Fluid Mechanics. 1985;158:153--175.

\bibitem{Jefferys1986}
Jefferys R, Whittaker T.
\newblock Latching control of an oscillating water column device with air
  compressibility.
\newblock In: Hydrodynamics of Ocean Wave-Energy Utilization. Springer; 1986.
  p. 281--291.

\bibitem{Taylor1919}
Taylor GI.
\newblock On the shapes of parachutes.
\newblock In: Batchelor GK, editor. The Scientific Papers of G. I. Taylor.
  Cambridge University Press; 1963. p. 26--37.
\newblock (Original work published 1919).

\bibitem{Harrison1970}
Harrison HB.
\newblock The analysis and behaviour of inflatable membrane dams under static
  loading.
\newblock Proceedings of the Institution of Civil Engineers.
  1970;45(4):661--676.

\bibitem{WAMIT7.2}
WAMIT.
\newblock Chestnut Hill, MA; 2016.
\newblock Version 7.2.
\newblock Available from: \url{http://www.wamit.com}.

\bibitem{FalcaoHenriques2014}
Falc{\~a}o AF, Henriques JC.
\newblock Model-prototype similarity of oscillating-water-column wave energy
  converters.
\newblock Int J Mar Energy. 2014;6:18--34.

\bibitem{Chaplin2012}
Chaplin JR, Heller V, Farley FJM, Hearn GE, Rainey RCT.
\newblock Laboratory testing the {A}naconda.
\newblock Phil Trans R Soc A. 2012;370(1959):403--424.

\bibitem{Beatty2015}
Beatty SJ, Hall M, Buckham BJ, Wild P, Bocking B.
\newblock Experimental and numerical comparisons of self-reacting point
  absorber wave energy converters in regular waves.
\newblock Ocean Engineering. 2015;104:370--386.

\bibitem{Pagitz2010}
Pagitz M, Pellegrino S.
\newblock Maximally stable lobed balloons.
\newblock Int Journal of Solids and Structures. 2010;47(11--12):1496--1507.

\bibitem{Falnes2015}
Falnes J, Kurniawan A.
\newblock Fundamental formulae for wave-energy conversion.
\newblock Royal Society open science. 2015;2(3):140305.

\bibitem{Budal1975}
Budal K, Falnes J.
\newblock A resonant point absorber of ocean-wave power.
\newblock Nature. 1975;256:478--479.
\newblock With Corrigendum in Nature, vol.~257, p.~626, 1975.

\bibitem{Newman1976}
Newman JN.
\newblock The interaction of stationary vessels with regular waves.
\newblock In: Eleventh Symposium on Naval Hydrodynamics. London; 1976. p.
  491--501.

\bibitem{Evans1976}
Evans DV.
\newblock A theory for wave power absorption by oscillating bodies.
\newblock Journal of Fluid Mechanics. 1976;77(1):1--25.

\bibitem{BudalFalnes1980}
Budal K, Falnes J.
\newblock Interacting point absorbers with controlled motion.
\newblock In: Count BM, editor. Power from Sea Waves. Academic Press; 1980. p.
  381--399.

\end{thebibliography}

\end{document}